\documentclass[10pt,conference]{IEEEtran}
\newcommand{\bx}{\ensuremath{{\mathbf x}}}
\newcommand{\bc}{\ensuremath{{\mathbf c}}}

\newcommand{\bR}{\ensuremath{{\mathbf R}}}

\newcommand{\br}{\ensuremath{{\mathbf r}}}

\newcommand{\blambda}{\mbox{\boldmath \ensuremath{\lambda}}}

\newcommand{\bmu}{\mbox{\boldmath \ensuremath{\mu}}}

\newcommand{\BEAS}{\begin{eqnarray*}}
\newcommand{\EEAS}{\end{eqnarray*}}
\newcommand{\BEA}{\begin{eqnarray}}
\newcommand{\EEA}{\end{eqnarray}}
\newcommand{\BEQ}{\begin{equation}}
\newcommand{\EEQ}{\end{equation}}
\newcommand{\BIT}{\begin{itemize}}
\newcommand{\EIT}{\end{itemize}}
\newcommand{\BNUM}{\begin{enumerate}}
\newcommand{\ENUM}{\end{enumerate}}

\newcommand{\BA}{\begin{array}}
\newcommand{\EA}{\end{array}}

\newcommand{\eg}{{\it e.g.}}
\newcommand{\ie}{{\it i.e.}}

\newcounter{exno}

{\begin{quote}}{\end{quote}}

\newcounter{oursection}

\newcounter{lecture}

\usepackage{graphics}
\begin{document}

\title{Distributed Algorithms for Optimal Rate-Reliability Tradeoff
in Networks}

\author{\authorblockN{Jang-Won Lee}
\authorblockA{Electrical Engineering Department\\
Princeton University\\
Email: janglee@princeton.edu} \and
\authorblockN{Mung Chiang}
\authorblockA{Electrical Engineering Department\\
Princeton University\\
Email: chiangm@princeton.edu} \and
\authorblockN{A. Robert Calderbank}
\authorblockA{Electrical Engineering Department\\
Princeton University\\
Email: calderbk@princeton.edu} }

\maketitle

\begin{abstract}
The current framework of network utility maximization for
distributed rate allocation assumes fixed channel code rates.
However, by adapting the physical layer channel coding, different
rate-reliability tradeoffs can be achieved on each link and for each
end user. Consider a network where each user has a utility function
that depends on both signal quality and data rate, and each link may
provide a `fatter' (`thinner') information `pipe' by allowing a
higher (lower) decoding error probability. We propose two
distributed, pricing-based algorithms to attain optimal
rate-reliability tradeoff, with an interpretation that each user
provides its willingness to pay for reliability to the network and
the network feeds back congestion prices to users. The proposed
algorithms converge to a tradeoff point between rate and
reliability, which is proved to be globally optimal for codes with
sufficiently large codeword lengths and user utilities with
sufficiently negative curvatures.
\end{abstract}

%%%%%%%%%%%%%%%%%%%%%%%%%%%%%%%%%%%%%%%%%%%%%%%%%%%%%%%%%%%%%%%

\section{Introduction}
\label{sec:introduction}

Since the publication of the seminal paper \cite{KMT98} by Kelly
\emph{et al.} in 1998, the framework of Network Utility Maximization
(NUM) has found many applications in network rate allocation
algorithms and Internet congestion control protocols, \eg, in
\cite{KS03,LA02,Low03,MW00}. Consider a communication network,
wireless or wired, with $L$ logical links, each with a fixed
capacity of $c_{l}$ bps, and $S$ sources (\ie, end users), each
transmitting at a source rate of $x_{s}$ bps. Each source $s$ emits
one flow, using a fixed set $L(s)$ of links in its path, and has a
utility function $U_{s}(x_{s})$. Each link $l$ is shared by a set
$S(l)$ of sources. NUM, in its basic version solved by a standard
distributed algorithm, is the following problem of maximizing the
total utility of the network $\sum_{s}U_{s}(x_{s})$, over the source
rates $\bx$, subject to linear flow constraints $\sum_{s\in S(l)}
x_{s} \leq c_{l}$ for all links $l$:
\begin{equation}
\label{eqn:utility1}
\begin{array}{ll}
\mbox{maximize}_{\bx} & \sum_{s}U_{s}(x_{s}) \\
\mbox{subject to}
& \sum_{s\in S(l)} x_{s} \leq c_{l}, \;\; \forall l, \\
& \bx\succeq 0.
\end{array}
\end{equation}

A major limitation of formulation (\ref{eqn:utility1}) is that
physical layer opportunities are entirely ignored. On links where
physical layer's adaptive resource allocation can change the
information `pipe' sizes, \eg, through power control and adaptive
coding, each link capacity $c_{l}$ is no longer a fixed constant but
a \emph{function} of resource allocation \cite{Chi05a}. For example,
changing the code rate on a link changes both the attainable
throughput and decoding error probability on the link. A larger
$c_{l}$ can be obtained on a link at the expense of lower decoding
reliability, which in turn lowers the end-to-end signal quality for
sources traversing the link and reduces user utilities at those
sources if utilities depend on both rate $x_{s}$ and reliability.
Dynamic adjustment of reliability provides an additional degree of
freedom for improving each user's utility as well as the system
efficiency. For example, if we allow lower decoding reliability on
more congested links and higher decoding reliability on less
congested links, we may be able to improve end-to-end joint
rate-reliability performance for all users. Such physical layer
opportunities are to be leveraged in both wireless systems and wired
networks such as DSL broadband access. However, the standard
distributed algorithm based on congestion prices (\eg,
\cite{KMT98,Low03}) cannot be applied since the concept of noisy
link and physical layer coding are absent from the basic NUM
formulation (\ref{eqn:utility1}).

In this paper, we exploit the tradeoff between information data rate
and signal reliability attained for each source, by controlling the
code rate, or equivalently, decoding error probability, for each
source's flow on each link and end-to-end information data rate of
each sources' flow. As in \cite{Chi05a}, we study a joint transport
and physical layer problem in the context of `layering as
optimization decomposition'. In contrast to \cite{Chi05a}, where
power control is considered at the physical layer, in this paper, we
consider reliability of each link through error control coding at
the physical layer. The new optimization formulation we propose in
general has coupled and nonconvex constraints. Two pricing-based
distributed algorithms proposed converge to the globally optimal
rate-reliability tradeoff under certain sufficient conditions. Two
approaches will be examined: the \emph{integrated dynamic
reliability policy}, where a link provides the same error
probability (\ie, the same code rate) to each of the sources whose
flows traverse it, and the \emph{differentiated dynamic reliability
policy}, where a link can provide a different error probability
(\ie, a different code rate) to each of the sources.

\section{System model}
\label{sec:system}

We assume that each source $s$ has a utility function
$U_s(x_s,R_s)$, where $x_s$ is an information data rate and $R_s$ is
reliability of received bits of an elastic traffic source $s$.
Similar to most models based on NUM, we assume that $U_{s}$ is a
smooth, increasing, and strictly concave function of $x_s$ and
$R_s$. Information data rate for each source $s$ may be constrained
within a range $[x_s^{min},x_s^{max}]$, and there may be a minimum
reliability requirement $R_s^{min}$. The reliability of source $s$
is defined as $R_s = 1-p_s^U$ where $p_s^U$ is an end-to-end error
probability for the flow from source $s$. Each link $l$ has its
maximum transmission capacity $C_l^{max}$. After link $l$ receives
the data of source $s$ from the upstream link, it first decodes it
to extract the information data and encodes it again with its own
code rate, $r_{l,s}$, where code rate is defined as the ratio of the
information data rate $x_{s}$ (bps) at the input of the encoder to
the transmission data rate $t_{l,s}$ (bps) at the output of the
encoder \cite{Gal68}. This allows the link to adjust both
transmission rate and error probability, since the transmission rate
$t_{l,s}$ of source $s$ at link $l$ can be defined as
\[
t_{l,s} = \frac{x_s}{r_{l,s}}
\]
and the error probability of source $s$ at link $l$ can be defined
as a function of  $r_{l,s}$ by
\[
p^L_{l,s} = E_l(r_{l,s}).
\]
This function $E_l(r_{l,s})$ rarely has a closed-form analytic
expression, and will be approximated through known bounds. The
end-to-end error probability for each source $s$ is $p_s^U = 1 -
\Pi_{l \in L(s)} (1-p_{l,s}^{L})= 1 - \Pi_{l \in L(s)}
(1-E_l(r_{l,s}))$. Assuming small decoding error probability (\ie,
$p_{l,s}^{L} \ll 1$), we can approximate the end-to-end error
probability of source $s$ as
\[
p_s^U \approx \sum_{l \in L(s)} p_{l,s}^{L} = \sum_{l \in L(s)}
E_l(r_{l,s}).
\]
Hence, the reliability of source $s$ can be written as
\[
R_s \approx 1-\sum_{l \in L(s)} E_l(r_{l,s}).
\]
Since each link $l$ has its transmission capacity $C_l^{max}$, the
sum of transmission rates of sources that are traversing each link
cannot exceed its transmission capacity, \ie,
\[
\sum_{s \in S(l)} t_{l,s} = \sum_{s \in S(l)} \frac{x_s}{r_{l,s}}
\leq C_l^{max},~ \forall l.
\]

\section{Integrated dynamic reliability policy}
\label{sec:integrated} We first investigate a more restrictive
policy where a link provides the \emph{same} code rate to each of
the sources whose flows traverse it, \ie,
\[
r_{l,s} = r_{l},~\forall s \in S(l),~ \forall l,
\]
and the NUM problem is formulated as
\begin{equation}
\label{eqn:int_primal}
\begin{array}{ll}
\mbox{maximize}_{\bx,\bR,\br} & \sum_{s} U_s(x_s,R_s)  \\
\mbox{subject to} & R_s \leq 1 -  \sum_{l \in L(s)}
            E_l(r_{l}), \;\; \forall s  \\
        &  \sum_{s \in S(l)} \frac{x_s}{r_{l}}
            \leq C_l^{max}, \;\; \forall l  \\
        & x_s^{min} \leq x_s \leq x_s^{max},\;\; \forall s \\
        & R_s^{min} \leq R_s \leq 1, \;\; \forall s \\
        & 0 \leq r_{l} \leq 1, \;\; \forall l.
\end{array}
\end{equation}

In order to derive a distributed algorithm to solve the above
problem and to prove convergence to global optimum, the critical
properties of separability and convexity of problem
(\ref{eqn:int_primal}) need to be carefully examined. Because of
the physical layer coding and rate-reliability tradeoff we
introduced into the problem formulation, these two properties no
longer trivially hold as in the basic NUM (\ref{eqn:utility1}).

Integrated policy naturally leads to a decomposition of problem
(\ref{eqn:int_primal}), since the second constraint can be written
as
\begin{equation}
\label{eqn:modified} \sum_{s \in S(l)} x_s \leq C_l^{max} r_{l}, ~
\forall l.
\end{equation}

The more complicated issue is convexity of function $E_l(r_{l})$. If
random coding based on $M$-ary binary coded signals is used, an
upper bound on the error probability is \cite{Pro89}:
\[
p_l < \frac{1}{2}2^{-N(R_0-r_l)},
\]
where $N$ is the block length and $R_0$ is the cutoff rate. Hence,
if we take
\[
E_l(r_l) = \frac{1}{2}2^{-N(R_0-r_l)},
\]
it is a convex function for given $N$ and $R_0$. A more general
approach is to use the random code ensemble error exponent
\cite{Gal68} that upper bounds the decoding error probability:
\[
p_l \leq \exp(-NE_{0}(r_l)),
\]
where $N$ is the codeword block length and $E_{0}(r_{l})$ is the
error exponent function. In general, $E_l(r_l)=\exp(-NE_{0}(r_l))$
may not be convex, even though it is known \cite{Gal68} that
$E_{0}(r_{l})$ is a convex function. The following lemma provides a
sufficient condition for convexity of $E_l(r_l)$.

\textbf{Lemma 1}. If the absolute value of the first derivatives of
$E_{0}(r_{l})$ is bounded away from 0 and the absolute value of the
second derivative of $E_{0}(r_{l})$ is upper bounded, then for a
large enough codeword block length $N$, $E_l(r_l)$ is convex.

\begin{proof}
Assume that there exist positive constants $\epsilon_1$ and
$\epsilon_2$ such that $\|\frac{d E_{0}(r_l)}{d r_l}\| \geq
\epsilon_1$ and $\left\|\frac{d^2 E_{0}(r_l)}{d r_l^2}\right\| \leq
\epsilon_2$.
\setlength{\arraycolsep}{2pt}
\begin{eqnarray*}
\frac{d^2 E_l(r_l)}{d r_l^2} &=& N \exp(-N E_{0}(r_l))(N(\frac{d
E_{0}(r_l)}{d r_l})^2 - \frac{d^2 E_{0}(r_l)}{d r_l^2}) \\
&\geq& N\exp(-NE_{0}(r_l))(N\epsilon_1^2 - \epsilon_2).
\end{eqnarray*}
\setlength{\arraycolsep}{5pt} Hence, $\frac{d^2 E_l(r_l)}{d r_l^2}
> 0$ for $N >
\frac{\epsilon_2}{\epsilon_1^2}$.
\end{proof}

Throughout this paper, we assume that the conditions in the above
lemma are true and the channel code is strong enough, \ie, $N$ is
large enough. In such cases, problem (\ref{eqn:int_primal}) is a
convex and separable optimization problem, which can be solved by
the following distributed algorithm where each source and each link
solve their own problems only with local information and limited
message passing. This distributed algorithm is derived by
decomposing the dual problem of (\ref{eqn:int_primal}) (not shown
explicitly in this summary), and we can interpret dual variables
$\lambda_{l}$ and $\mu_s$ as the price per unit rate to use link $l$
and the price per unit reliability that the source $s$ must pay to
the network, respectively.
%\begin{figure}[t!]
%\begin{center}
%\resizebox{6cm}{!}{\includegraphics{int_diagram.eps}}
%\caption{Diagram for the distributed algorithm of the integrated
%dynamic reliability policy.} \label{fig:int_diagram}
%\end{center}
%\end{figure}

\textbf{Algorithm 1.}

In each iteration $t$, by solving the following problem
(\ref{eqn:int_source})\footnote{Optimization problems
(\ref{eqn:int_source}) and (\ref{eqn:int_link}) (also problems
(\ref{eqn:diff_source}) and (\ref{eqn:diff_link}) in Algorithm 2)
are convex optimization in only one or two variables with simple
range constraints. Hence, they can be easily solved by using
standard algorithms such as a gradient projection algorithm.} over
$(x_{s},R_{s})$, each source $s$ determines its information data
rate and desired reliability (\ie, $x_s(t)$ and $R_s(t)$) that
maximize its net utility based on the prices
$(\lambda^{s}(t),\mu_{s}(t))$ in the current iteration. Furthermore,
by price update equation (\ref{eqn:int_price_source}), the source
offers price $\mu_s(t+1)$ per unit reliability for the next
iteration.

\textbf{Source problem and reliability price update at source $s$:}
\begin{equation}
\label{eqn:int_source}
\begin{array}{ll}
\mbox{maximize}_{x_s,R_s} &U_s(x_s, R_s)  - \lambda^{s}(t) x_s -\mu_s(t) R_s \\
\mbox{subject to} & x_s^{min} \leq x_s \leq x_s^{max} \\
                        & R_s^{min} \leq R_s \leq 1,
\end{array}
\end{equation}
where $\lambda^{s}(t) = \sum_{l \in L(s)} \lambda_l(t)$.
\setlength{\arraycolsep}{2pt}
\begin{eqnarray}
\label{eqn:int_price_source} \mu_s(t+1) &=& [ \mu_{s}(t) - \beta(t)
( R^s(t)-R_s(t))]^+, ~\forall s,
\end{eqnarray}
\setlength{\arraycolsep}{5pt} where $R^s(t) = 1 - \sum_{l \in L(s)}
E_l(r_l(t))$ and $\beta(t)$ is step size, which can be taken as
$\beta_{0}/t$ for some $\beta_{0}>0$.

Concurrently in each iteration $t$, by solving problem
(\ref{eqn:int_link}) over $r_{l}$, each link $l$ determines  its
code rate (\ie, $r_l(t)$) that maximizes the `net revenue' of the
network based on the prices in the current iteration. In addition,
by price update equation (\ref{eqn:int_price_link}), the link
adjusts its congestion price $\lambda_l(t+1)$ per unit rate for the
next iteration.

\textbf{Link problem and congestion price update at link $l$:}
\begin{equation}
\label{eqn:int_link}
\begin{array}{ll}
\mbox{maximize}_{r_l}&  \lambda_l(t)  r_l C_l^{max} -\mu^l(t) E_l(r_l) \\
\mbox{subject to} & 0 \leq r_l \leq 1,
\end{array}
\end{equation}
where $\mu^l(t) = \sum_{s \in S(l)}\mu_s(t)$.
\setlength{\arraycolsep}{2pt}
\begin{eqnarray}
\label{eqn:int_price_link} \lambda_{l}(t+1)&=&
[\lambda_{l}(t)-\beta(t) (r_l(t) C_l^{max}  - x^l(t))]^+,
     ~\forall l,
\end{eqnarray}
\setlength{\arraycolsep}{2pt} where $x^l(t) = \sum_{s \in S(l)}
x_s(t)$.

In the above Algorithm 1, source $s$ needs to know $\lambda^s(t)$,
the sum of $\lambda_l(t)$'s of links that are along its path $L(s)$.
This can be obtained by the notification from the links \eg, through
acknowledgment packets. To carry out price update
(\ref{eqn:int_price_source}), the source needs to know the sum of
error probabilities of the links that are along its path (\ie, its
own reliability that is provided by the network, $R^s(t)$). This can
be obtained by the notification from the links that determines the
code rate for the source (by solving problem (\ref{eqn:int_link}))
or obtained by the notification from the destination that measures
the end-to-end reliability. To solve the link problem
(\ref{eqn:int_link}), each link $l$ needs to know $\mu^l(t)$, the
sum of $\mu_s(t)$'s from sources $s\in S(l)$ using this link $l$.
This can be obtained by the notification from these sources. To
carry out update (\ref{eqn:int_price_link}), the link needs to know
the aggregate information data rate of the sources that are using it
(\ie, $x^l(t)$). This can be obtained by measuring it by the link
itself.

With dual decomposition and Lemma 1, the following result can be
proved using standard techniques in distributed gradient algorithm's
convergence proof:

\textbf{Theorem 1}. By Algorithm 1, dual variables $\blambda(t)$ and
$\bmu(t)$ converge to the optimal dual solutions $\blambda^{*}$ and
$\bmu^{*}$ and the corresponding primal variables $\bx^*$, $\bR^*$,
and $\br^*$ are the globally optimal primal solutions of
(\ref{eqn:int_primal}).

\section{Differentiated dynamic reliability policy}
\label{sec:differentiated} In this policy, a link may provide a
\emph{different} code rate to each of the sources traversing it.
An example of practical code constructions that enable such a
flexibility is the coding technique with embedded diversity
recently proposed in \cite{CDA04a}, which allows data streams with
different rate-reliability tradeoffs be embedded within each
other.

The problem formulation in (\ref{eqn:int_primal}) needs to be
generalized to the following:
\begin{equation}
\label{eqn:diff_primal}
\begin{array}{ll}
\mbox{maximize}_{\bx,\bR,\br}&  \sum_{s} U_s(x_s,R_s) \\
\mbox{subject to}
        & R_s \leq 1 - \sum_{l \in L(s)}
            E_l(r_{l,s}), \;\; \forall s \\
        &  \sum_{s \in S(l)} \frac{x_s}{r_{l,s}}
            \leq C_l^{max}, \;\; \forall l \\
        & x_s^{min} \leq x_s \leq x_s^{max},\;\; \forall s\\
        & R_s^{min} \leq R_s \leq 1, \;\; \forall s \\
        & 0 \leq r_{l,s} \leq 1, \;\; \forall l, ~s \in S(l).
    \end{array}
\end{equation}
The objective function and constraints of problem
(\ref{eqn:diff_primal}) are the same as those of problem
(\ref{eqn:int_primal}) except that we have $r_{l,s}$ here instead of
$r_l$. Due to this critical difference, problem
(\ref{eqn:diff_primal}) in general is neither a convex problem nor a
separable one (since we may not modify the second constraint in this
problem as in (\ref{eqn:modified})). To resolve this issue, we
transform problem (\ref{eqn:diff_primal}) by first introducing the
auxiliary variables $c_{l,s}$, which can be interpreted as the
allocated transmission capacity to source $s$ at link $l$. Then, the
above problem can be reformulated as
\begin{equation}
\label{eqn:diff_primal1}
\begin{array}{ll}
\mbox{mazimize}_{\bx,\bR,\br} &  \sum_{s} U_s(x_s, R_s) \\
\mbox{subject to}
        & R_s \leq 1-\sum_{l \in L(s)}
            E_l(r_{l,s}), \;\; \forall s\\
        & \frac{x_s}{r_{l,s}} \leq c_{l,s}, \;\; \forall l,
            ~ s \in S(l)\\
        & \sum_{s \in S(l)} c_{l,s}
            \leq C_l^{max}, \;\; \forall l \\
        & x_s^{min} \leq x_s \leq x_s^{max},\;\;\forall s \\
        & R_s^{min} \leq R_s \leq 1, \;\; \forall s \\
        & 0 \leq r_{l,s} \leq 1, \;\; \forall l,~ s \in S(l)\\
        & 0 \leq c_{l,s} \leq C_l^{max}, \;\; \forall l,~ s \in
        S(l).
    \end{array}
\end{equation}
The second constraint in problem (\ref{eqn:diff_primal}) is now
decomposed into two constraints in problem (\ref{eqn:diff_primal1}):
the second and third constraints. Here, the second constraint
implies that the transmission data rate of each source at each link
must be smaller than or equal to its allocated transmission capacity
at the link, and the third constraint implies that the aggregate
allocated transmission capacity to the sources at each link must be
smaller than or equal to its maximum transmission capacity. In this
formulation, each link explicitly allocates a transmission capacity
to each of its sources. We can easily show that problem
(\ref{eqn:diff_primal1}) is equivalent to problem
(\ref{eqn:diff_primal}), since at optimality, the second constraint
in problem (\ref{eqn:diff_primal1}) is satisfied with the equality.

The next step of problem transformation is to take a log change of
variable at the second constraint in problem
(\ref{eqn:diff_primal1}) and set $x'_s = \log{x_s}$ (\ie, $x_s =
e^{x'_s}$). This reformulation turns problem (\ref{eqn:diff_primal})
into the following equivalent problem:
\begin{equation}
\label{eqn:diff_primal2}
\begin{array}{ll}
\mbox{maximize}_{\bx',\bR,\br,\bc} & \sum_{s} U'_s(x'_s, R_s) \\
\mbox{subject to}
        & R_s \leq 1-\sum_{l \in L(s)}
            E_l(r_{l,s}), \;\; \forall s\\
        & x'_s - \log{r_{l,s}} \leq \log{c_{l,s}},
            \;\; \forall l,~ s \in S(l)\\
        &  \sum_{s \in S(l)}
            c_{l,s} \leq C_l^{max}, \;\; \forall l\\
        & x_s^{'min} \leq x'_s \leq x_s^{'max},\;\; \forall s \\
        & R_s^{min} \leq R_s \leq 1, \;\; \forall s \\
        & 0 \leq r_{l,s} \leq 1, \;\; \forall l,~ s \in S(l)\\
        & 0 \leq c_{l,s} \leq C_l^{max}, \;\; \forall l,~ i \in S(l),
    \end{array}
\end{equation}
where $U'_s(x'_s, R_s) = U_s(e^{x'_s}, R_s)$ and $x_s^{'min} =
\log{x_s^{min}}$ and $x_s^{'max} = \log{x_s^{max}}$.

Note that problem (\ref{eqn:diff_primal2}) is separable but still
may not be a convex optimization problem since the objective
$U'_s(x'_s, R_s)$ may not be a concave function, even though
$U_s(x_s,R_s)$ is a concave function. The following lemma provides a
sufficient condition for its concavity. For notational simplicity,
assume that $U_s(x_s,R_s)$ is additive in each variable (the general
case is an easy extension), $U_s(x_s,R_s) = U_s^x(x_s)+U_s^R(R_s)$
(\ie, $U'_s(x'_s,R_s)=U^{x'}_i(x'_s) + U_s^R(R_s)$), and
$U_s^R(R_s)$ be a strictly concave function of $R_s$. Define
\[
g_s(x_s)=\frac{d^2 U_s^x(x_s)}{dx_s^2} x_s +
        \frac{dU_s^x(x_s)}{dx_s}.
\]

\textbf{Lemma 2}. If $g_s(x_s) \leq 0$, $U_s^{x'}(x'_s)$ is a
concave function of $x'_s$ and $U'_s(x'_s,R_s)$ is a  concave
function of $x'_s$ and $R_s$.

\begin{proof}
Since $x_s = e^{x'_s}$,
\begin{eqnarray*}
\frac{d^2 U_s^{x'}(x'_s)}{dx_s^{'2}} &=& \frac{d^2
U_s^{x}(x_s)}{dx_s^2} \left(\frac{dx_s}{dx'_s}\right)^2 +
\frac{dU_s^x(x_s)}{dx_s} \frac{d^2 x_s}{dx_s^{'2}} \\
&=& e^{x'_s}\left(\frac{d^2 U_s^{x}(x_s)}{dx_s^2} e^{x'_s} +
\frac{dU_s^x(x_s)}{dx_s}\right)\\
%&=& e^{x'_s}\left(\frac{d^2 U_s^{x}(x_s)}{dx_s^2}x_s +
%\frac{dU_s^x(x_s)}{dx_s}\right)\\
&=& e^{x'_s}g_s(x_s).
\end{eqnarray*}
Hence, if $g_s(x_s) \leq 0$, $U_s^{x'}(x'_s)$ is a concave function
of $x'_s$
\end{proof}

The condition of $g_{s}(x_{s})\leq 0$ is equivalent to:
\[
\frac{d^2 U_s^x(x_s)}{dx_s^2} \leq -\frac{dU^x_s(x_s)}{x_{s}dx_s}.
\]
Since utility functions are increasing, $\frac{dU^x_s(x_s)}{dx_s}$
is a positive number. The above inequality states that the utility
function needs to be not just concave (\ie, $\frac{d^2
U^x_s(x_s)}{dx_s^2} \leq 0$), but with a curvature that is bounded
away from 0 by as much as $\frac{dU^x_s(x_s)}{x_{s}dx_s}$, \ie, the
user utility function must be `elastic' enough.

For example, consider the often-used utility functions
\cite{MW00}:
\[
U_s^{\alpha}(x_s) = \left \{ \begin{array}{ll}
                \log{x_s}, & \mbox{if } \alpha = 1 \\
                (1-\alpha)^{-1}x_s^{1-\alpha}, & \mbox{otherwise}
                \end{array}
                \right..
\]
For this family of utility functions parameterized by $\alpha$,
Lemma 2 shows that problem (\ref{eqn:diff_primal2}) becomes a
convex optimization problem if $\alpha \geq 1$.

After the above procedures of problem transformations, we are
ready to provide a distributed algorithm based on dual
decomposition to solve problem (\ref{eqn:diff_primal2}), and to
prove the performance guarantee on global optimality provided that
the conditions in Lemmas 1 and 2 hold.

\textbf{Algorithm 2.}

Equations (\ref{eqn:int_source})-(\ref{eqn:int_price_link}) are now
replaced by (\ref{eqn:diff_source})-(\ref{eqn:price_link}). In
contrast to Algorithm 1 that does not keep per-flow information on
the links, in the more complex Algorithm 2, each link differentiates
each of the flows through it by providing a different code rate
$r_{l,s}$, a different congestion price $\lambda_{l,s}$, and an
explicit capacity allocation $c_{l,s}$. In addition, the congestion
price $\lambda_{l,s}$ is determined based on the allocated capacity
$c_{l,s}$ and the transmission rate $x_s/r_{l,s}$ of each individual
source that uses link $l$ in Algorithm 2, while the congestion price
$\lambda_l$ is determined based on the aggregate transmission rate
of sources that use link $l$ and the transmission capacity of the
link in Algorithm 1.
%In each iteration $t$, by solving (\ref{eqn:diff_source}) over
%$(x'_s, R_s)$, each source $s$ determines its information data rate
%and desired reliability  (\ie, $x'_s(t))$ or equivalently, $x_s(t)
%= e^{x'_s(t)}$, and $R_s(t)$) that maximize its net utility based on
%the prices in the current iteration. Furthermore, by price update
%equation (\ref{eqn:price_source}), the source adjusts its offered
%price per unit reliability for the next iteration.

\textbf{Source problem and reliability price update at source $s$:}
\begin{equation}
\begin{array}{ll}
\mbox{mazimize}_{x'_s,R_s} & U_s(x'_s, R_s)- \lambda^s(t) x'_s -\mu_i(t) R_s\\
\mbox{subject to}& x^{'min}_s \leq x'_s \leq x^{'max}_s \\
                            & R_s^{min} \leq R_s \leq 1,
\end{array}
\label{eqn:diff_source}
\end{equation}
where $\lambda^s(t) = \sum_{l \in L(s)} \lambda_{l,s}(t)$.
\setlength{\arraycolsep}{1pt}
\begin{eqnarray}
\label{eqn:price_source} \mu_s(t+1) &=&[ \mu_s(t) - \beta(t) (R^s(t)
-R_s(t))]^+, ~\forall s,
\end{eqnarray}
\setlength{\arraycolsep}{2pt} where $R^s(\blambda(t), \bmu(t)) = 1-
\sum_{l \in L(s)} E_{l}(r_{l,s}(\blambda(t), \bmu(t))$.

%Concurrently in each iteration $t$, by solving problem
%(\ref{eqn:diff_link}) over $(c_{l,s},r_{l,s})$, $\forall s \in
%S(l)$, each link $l$ determines the allocated transmission capacity
%and the code rate of each of its sources (\ie, $c_{l,s}(t)$ and
%$r_{l,s}(t)$) that maximize the `net revenue' of the network based
%on the prices in the current iteration. In addition, by  price
%update equation (\ref{eqn:price_link}), the link adjusts its
%congestion price per unit rate of source $s$ for the next iteration.

\textbf{Link problem and congestion price update at link $l$:}
\begin{equation}
\label{eqn:diff_link}
\begin{array}{ll}
\mbox{maximize}_{r_{l,s},c_{l,s},~s \in S(l)} &  \sum_{s \in S(l)}\{
\lambda_{l,s}(t)
(\log{c_{l,s}}+\log{r_{l,s}})\\
& - \mu_s(t) E_l(r_{l,s}) \}\\
\mbox{subject to} & \sum_{s \in S(l)} c_{l,s} \leq C_l^{max}  \\
&   0 \leq c_{l,s} \leq C_l^{max}, \;\; s \in S(l) \\
&   0 \leq r_{l,s} \leq 1, \;\; s \in S(l).
\end{array}
\end{equation}
\setlength{\arraycolsep}{1pt}
\begin{eqnarray}
\label{eqn:price_link}  \lambda_{l,s}(t+1) &=& [\lambda_{l,s}(t) -
\beta(t) ( \log{c_{l,s}(t)}
+ \log{r_{l,s}(t)} - x'_s(t))]^+ \nonumber \\
%&=& [\lambda_{l,s}(t) - \beta(t) ( \log{c_{l,s}(t)}+
%\log{r_{l,s}(t)} - \log{x_s(t)})]^+, \nonumber \\
& &~\forall l, s \in S(l).
\end{eqnarray}
\setlength{\arraycolsep}{5pt}

%\vspace{-0.3cm}
\textbf{Theorem 2}. By Algorithm 2, dual variables $\blambda(t)$ and
$\bmu(t)$ converge to the optimal dual solutions $\blambda^{*}$ and
$\bmu^{*}$ and the corresponding primal variables $\bx^{'*}$,
$\bR^*$, $\bc^*$, and $\br^*$ are the globally optimal primal
solutions of problem (\ref{eqn:diff_primal2}).

Extensions of Algorithms 1 and 2 to asynchronous version with
constant step size can be carried out similar to those in
\cite{Chi05a}.

\section{Numerical examples}
\label{sec:numerical}
\begin{figure}[h!]
\begin{center}
\resizebox{5.cm}{!}{\includegraphics{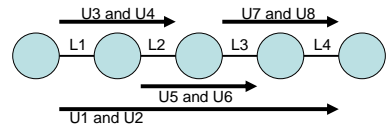}} \caption{Network
topology and flow routes for rate-reliability tradeoff example.}
\label{fig:tradeoffnet}
\end{center}
\end{figure}
We present numerical examples for a simple network shown in Figure
\ref{fig:tradeoffnet}. Utility function $U_{s}(x_{s},R_{s})$ for
user $s$ has the following standard form of concave utility:
%shifted such that
%$U_{s}(x_{s}^{min},R_{s}^{min})=0$, and with utility on rate and
%utility on reliability summed up with a given weight $a_{s}$:
%\[
%\frac{a_{s}}{1-\alpha}[x_{s}^{1-\alpha}-x_{s}^{min(1-\alpha)}]
%+\frac{1-a_{s}}{1-\alpha}[R_{s}^{(1-\alpha)}-R_{s}^{min(1-\alpha)}],
%\]
\begin{eqnarray*}
U_{s}(x_s,R_s) &=&
a_{s}\frac{x_{s}^{1-\alpha}-x_{s}^{min(1-\alpha)}}
{x_{s}^{max(1-\alpha)}-x_{s}^{min(1-\alpha)}}
\\
& &
+(1-a_{s})\frac{R_{s}^{(1-\alpha)}-R_{s}^{min(1-\alpha)}}
{R_{s}^{max(1-\alpha)}-R_{s}^{min(1-\alpha)}}.
\end{eqnarray*}
where $a_{s}$ is a constant ($0 \leq a_s \leq 1$) that determines
the relative weight between rate and reliability. The decoding error
probability on each link $l$ is assumed to be the following form:
$p_l^L = \frac{1}{2}\exp(-N(1-r_{l}))$ where $N$ is the channel code
block length and $r_{l}$ the code rate for link $l$. We set $\alpha
= 1.1$, $x_i^{min} = 0.1$ (Mbps), $x_i^{max} = 2$ (Mbps), $C_l^{max}
= 2$ (Mbps), $R_i^{max} = 1$, and $R_i^{min} = 0.9$.

We trace the globally optimal tradeoff curve between rate and
reliability, and also compare the total network utility achieved by
the following three policies:
\begin{itemize}
\item Static reliability
%: a fixed error probability 0.01 at each link.
\item Integrated dynamic reliability
%: the same adjustable error probability
%    to each of its users at each link.
\item Differentiated dynamic reliability.
%: a possibly different
%error probability to each of its users at each link.
\end{itemize}
In the static scheme, each link provides a fixed error probability
0.025 and rate is allocated to each user by solving the basic NUM
problem (\ref{eqn:utility1}).

\begin{figure}[t]
\begin{center}
\resizebox{5.cm}{!}{\includegraphics{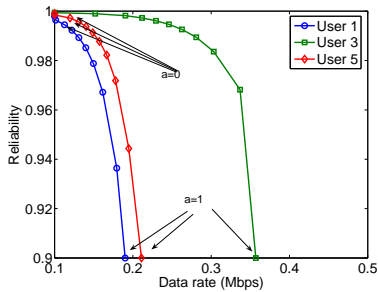}} \caption{Globally
optimal rate-reliability tradeoff of 3 of the end users, using the
integrated dynamic reliability policy (Algorithm 1).}
\label{fig:user}
\end{center}
\end{figure}

We first investigate the case where all the users have the same
$a_{s}=a$, and vary the value of $a$ from 0 to 1 in step size of
0.1. The resulting tradeoff curve obtained by Algorithm 1, which
is globally optimal, is shown in Figure \ref{fig:user}. As
expected, a larger $a$ (\ie, utility from rate given a heavier
weight) leads to a higher rate at the expense of lower
reliability. The more congested links a user's flow passes
through, the steeper the tradeoff curve becomes. For each user,
the area to the left and below of the tradeoff curve is the
\emph{achievable region}, and the area to the right and above of
the tradeoff curve is the \emph{infeasible region}. It is
impossible to operate in the infeasible region and inferior to
operate in the interior of the achievable region. Operating on the
boundary of the achievable region, \ie, the Pareto optimal
tradeoff curve, is the best.

\begin{figure}[t]
\begin{center}
\resizebox{5.cm}{!}{\includegraphics{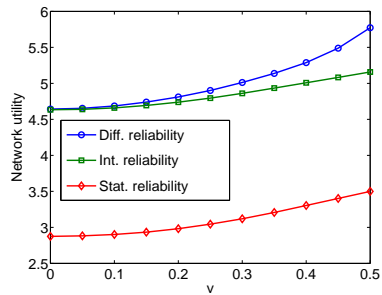}}
\caption{Comparison of the achieved network utility attained by each
policy.}
%differentiated dynamic policy (proposed Algorithm 2),
%integrated dynamic policy (proposed Algorithm 1), and static policy
%(current practice).}
\label{fig:diff_user}
\end{center}
\end{figure}

We then give different weights $a_{s}$ to the eight users:
\[
a_{s} = \left \{ \begin{array}{ll} 0.5 - v,&\mbox{if $s$ is an odd number}\\
                                    0.5+v,&\mbox{if $s$ is an even number}
                    \end{array} \right.
\]
and vary $v$ from 0 to 0.5 in step size of 0.05. Figure
\ref{fig:diff_user} shows the relative performance in terms of
network utility achieved by the three policies as $v$ changes. The
performance of Algorithms 1 and 2 that take into account
rate-reliability tradeoff is significantly better than the standard
distributed algorithm for the basic NUM (\ref{eqn:utility1}) that
ignores the possibility of jointly optimizing rate and reliability.

%%%%%%%%%%%%%%%%%%%%%%%%%%%%%%%%%%%%%%%%%%%%%%%%%%%%%%%

\section{Concluding Remarks}
\label{sec:conclusion}

Motivated by needs from the application layer and possibilities at
the physical layer, this paper removes the rate-dependency
assumption on utility functions in the current NUM formulation and
allows the physical layer adaptive channel coding to tradeoff rate
and reliability on each link and for all the sources. We present two
distributed algorithms for two possible formulations of the problem,
and provide the sufficient conditions on codeword length and utility
curvature under which convergence to the globally optimal
rate-reliability tradeoff can be proved.
%The information-theoretic concept of error exponents is used in constraint models.
In addition to link-updated congestion prices for distributed rate
control, we also introduce source-updated signal quality prices for
distributed reliability control. The tradeoff between packet drop
probability and traffic load may be investigated similar to that
between decoding error probability and transmission rate examined in
this paper.

%We have assumed that each relay node decodes and re-encodes. For
%wireless multihop networks, a simple strategy of forwarding without
%decoding, or a multiuser decoding strategy that does not treat
%interference as noise, may be more efficient. It is an interesting
%next step to characterize the benefits of such physical layer
%possibilities at the end user utility level by studying distributed
%algorithms that would solve for the global optimum of the
%corresponding NUM problem.

\end{document}